\documentstyle[11pt,aaspp4]{article}

\slugcomment{}

\lefthead{Stiavelli et al.}
\righthead{Lyman-$\alpha$ emitters at $z\simeq2.4$}

\begin{document}

\title{Lyman-$\alpha$ emitters with red colors at $z\simeq2.4$}

\author{Massimo Stiavelli\altaffilmark{1}, Claudia Scarlata
\altaffilmark{2}} \affil{Space Telescope Science Institute, 3700 San
Martin Dr., 21218 MD, U.S.A.; mstiavel@stsci.edu, scarlata@stsci.edu}
\author{Nino Panagia \altaffilmark{3}}\affil{Space Telescope Science Institute, 3700 San Martin Dr., 21218 MD, U.S.A.; panagia@stsci.edu}
\author{Tommaso Treu}\affil{California Institute of Technology, MS 105-24, Pasadena, CA 91125, U.S.A.; tt@astro.caltech.edu}
\author{Giuseppe Bertin}\affil{Universit{\`a} degli Studi di Milano, via Celoria 16, I20133 Milano, Italy; Giuseppe.Bertin@mi.infn.it}
\author{Francesco Bertola}\affil{Universit{\`a} degli Studi di Padova, vicolo dell'Osservatorio 5, I35122, Padova, Italy;
bertola@pd.astro.it}

\altaffiltext{1}{Visiting Astronomer, Canada-France-Hawaii Telescope,
which is operated by the National Research Council of Canada, the Centre
Nationale de la Recherche Scientifique of France, and the University of Hawaii}
\altaffiltext{2}{also Universit{\`a} degli Studi di Padova, Italy}
\altaffiltext{3}{On assignment from the Space Science Dept. of ESA}

\begin{abstract}

We have carried out a search for Lyman-$\alpha$ emission from galaxies
at z$\simeq$2.4 over a field of 1200 $\sq '$ using the CFH12K camera
at the CFHT and a custom medium band filter. The search has uncovered
58 candidates, corresponding to a completeness-corrected source
density of $\sim$0.07 $\sq '$ $^{-1}$. Our sources have red colors
($\overline{B-I}$ $\simeq$ 1.8) which imply either that a large
fraction of the light is highly reddened and we are detecting
Lyman-$\alpha$ through special lines of sight, or that these objects
contain an underlying older stellar population. While for each
individual object we cannot discriminate between these alternatives,
we conclude that most of the objects actually contain an older
component because the star formation rates inferred from the picture
based on reddening, applied to all candidates, would imply an
exceedingly high star formation rate, {\it i.e.}  more than two orders
of magnitude above the peak cosmic star formation rate (e.g. Lilly et
al. 1996).

\end{abstract}



\section{Introduction}

In recent years, the study of the global star formation rate of
the Universe has become a very popular subject after it was introduced
and developed as a diagnostic technique by a number of seminal papers
(\cite{PF95}, \cite{Lilly}, \cite{MP}.) Several authors have noted
some limitations of the original approach based on the HDFN data
(\cite{MP}) especially for two reasons:

{\it i)} Estimating the star formation rate (hereafter SFR) from the
UV flux might underestimate the true SFR because of the
effects of dust extinction (\cite{MHC}, \cite{Calzetti}, \cite{Peietal})

{\it ii)} The small area of the HDF may suffer from cosmic variance,
especially because the brightest, relatively rare, sources may be
seriously affected by small number statistics and clustering
(\cite{Steidel}, \cite{Steidelb}).

Even disregarding these limitations, the cosmic star formation
history provides only partial information about how
individual galaxies evolve. A low mean metallicity of
the Universe at some redshift might imply either that all
galaxies were in place and metal poor at that redshift or that only a
few, possibly metal rich, galaxies had formed early on. Indeed,
evidence for the latter interpretation may be found in the scatter in
metallicity of damped Lyman-$\alpha$ systems and in the relatively
high metallicity of the $z=0.5-1$ CFRS galaxies (\cite{LillyMarci}).

A clear answer to these questions could be found by simultaneously
measuring star formation rate, metallicity, and dust content for a
statistically significant sample of high redshift ($z\geq2$)
galaxies. A reliable measurement of the gas physical properties
requires access to the rest frame optical spectrum and in particular
to the Balmer lines, and to the [NII]$\lambda\lambda 6548, 6583$,
[OIII]$\lambda\lambda 4959,5007$ and [OII]$\lambda\lambda 3727, 3729$
doublets (e.g., \cite{KenniLiege}; \cite{StiavLiege}).  This has been
attempted only for a handful of objects (\cite{Kobulni}, see also
\cite{Moorwood}, \cite{Pettini2001}). The $z \simeq 2.4$ redshift
window is ideal for this kind of measurements since at this redshift
all of these lines fall in the Near-IR atmospheric windows. At the
same time $z \simeq2.4$ is larger than that of the peak in
the global star formation rate as claimed by Madau et al. (1996). For
this reason, we have started a search for Lyman-$\alpha$ emitters with
the CFH12K camera at the CFHT using a custom medium band filter
centered at $4158$ \AA. The area covered is 1200 $\sq '$. The survey
is briefly described in Section 2. Our sample of bona fide
Lyman-$\alpha$ emitters is described and discussed in Section
3. Section 4 contains our general discussion and conclusions.

\section{The survey}

\label{sec:survey}

The CFH12K camera at the CFHT is a mosaic of 12 MIT Lincoln Lab CCDs
each 2K by 4K pixels. The pixel size is 0.2'', yielding a total field
of view of 42 by 30 $\sq '$. We carried out our observations in July
27-August 1, 2000. The main filter used was a custom medium band
interference filter with central wavelength $4158$ \AA\ and width
$174$ \AA\ aimed at detecting Lyman-$\alpha$ at z=2.422$\pm$0.072. The
filter peak throughput is 65 \%. We obtained images for a total
exposure time of 19,800s on a field offset by $\sim$120 arcsec from
the QSO DMS 2139-0356 at $z\simeq 2.36$.  For the same field, we also
obtained exposures for 3000s in I band and 6600s in B band. After
standard reduction steps, the images were combined using a standard
CR-rejection algorithm (IRAF's {\it imcombine}). The photometric
calibration was based on the nominal zero-points for the camera and
checked by observing standard stars in the Landolt field SA 107
(\cite{Landolt}).  All magnitudes and fluxes were also corrected for
galactic extinction (E(B-V) = 0.055). A more detailed description of
the data reduction and analysis is presented separately (Scarlata et
al. 2001).

Once the images were reduced and combined, they were searched for
objects using the SExtractor package (\cite{SEx}). We verified our
object finding algorithm by comparing the B band number counts with
those of Metcalfe et al. (2001) and found good agreement down to
B$\simeq$25. The photometric accuracy of our procedure was tested extensively
by introducing artificial stars on our images and recovering them. The
same simulations provided us also with the completeness correction as
a function of flux.


Continuum subtraction was carried out by deriving the histogram of
Lyman-$\alpha$ to B band flux ratios for all sources in each chip
(excluding saturated stars).  From the histogram we
derived the median flux ratio and the FWHM of the histogram which we
use as a measure of the intrinsic cosmic dispersion in the ratio. Our
selection criterion requires the line excess to be both statistically
significant and larger than the cosmic dispersion. A source was
considered a candidate if the flux in the Lyman-$\alpha$ medium band
exceeded both: {\it i)} the continuum contribution - estimated with
the Lyman-$\alpha$ over continuum ratio - by more than 3-$\sigma$ and
{\it ii)} the value corresponding to the ratio plus one histogram
FWHM. The exclusion of sources with ratios within one histogram FWHM
eliminates high S/N sources within the cosmic scatter. The validity of
this procedure was checked by estimating the expected continuum
contribution in the Lyman-$\alpha$ medium band filter by using
spectral synthesis models (\cite{BC93}). For the selection of our
sample, we adopted the observed ratio rather than the theoretical
values because the former is robust with respect to uncertainties in
the filter effective wavelengths and in the exact QE shape of the
various CCDs and calibration uncertainties. Our results do not
critically depend on this choice. Our 3-$\sigma$ line flux limit for a
source without continuum is $\sim$2$\times$10$^{-16}$ erg s$^{-1}$
cm$^{-2}$.  A total of 89 sources were found with significant
emission in the medium band filter.  No overdensity was found
associated with the QSO DMS 2139-0356.

The most serious contamination of our sample is expected to be due to
[OII]$\lambda\lambda$3726, 3729 emitters at $z\simeq 0.11$. In order
to remove this contamination we considered the distribution of local
[OII] emitters (\cite{Jansen}) in the equivalent width
(EW)-M$_B$-(B-R) space and compared it to our sources (Figure 1). We
assumed that they were {\it all} located at z$\simeq$0.11 and compared
their properties with those of the local [OII] emitters (moved to
z=0.11). The B-R colors of our sources were obtained from the measured
B-I by appling a K-correction for $z\simeq 0.11$ derived from galaxy
templates and spectral synthesis models.  The comparison shows that
the local [OII] emitters are on average brighter than our
candidates. This is may be partly due to the magnitude limit of the
local sample but is also due to the fact that our sources would be
very faint (and very compact) if at low redshift. In addition, the
local emitters show trends of increasing EW with decreasing luminosity
(top left panel in Figure 1), increasing EW with bluer colors (top
right panel), and bluer colors with decreasing luminosity (bottom left
panel). We determined best fits to these relations and produced a
delta-delta plot by subtracting from the Log EW and B-R color of each
object the mean value corresponding to its absolute magnitude. In the
delta-delta plot we thus identify the bona fide Lyman-$\alpha$
emitters as those objects that lie outside the locus of the local
[OII] objects. Such an approach is pessimistic since it assumes that
all those objects that are along the mean relations observed for [OII]
emitters at low redshift are indeed such. A total of 31 sources are
thus rejected.  We should note that the Jansen et al. (2000) sample is
a fair representation of local galaxies only down to absolute
magnitude M$_B=-14$. One might imagine a population of fainter
starbursting dwarfs with intrinsically much higher [OII] EW
contaminating our sample of bona fide candidates. We think that this
is highly unlikely since: {\it i)} all local group dwarfs fainter than
M$_B=-14$ have little or no line emission, {\it ii)} even local
starburst and blue compact galaxies (see, e.g., \cite{mcquade}) do not
show [OII] EW in excess of 100. Thus, we will assume the validity of
our rejection and base our analysis on the remaining 58 objects,
implying a density of 0.07$\pm$0.01 sources \sq ' $^{-1}$ above $2
\times 10^{-16}$ erg cm$^{-2}$ s$^{-1}$. This density estimate has
been corrected for completeness and (statistically) for the number
($\sim$6) of 3-$\sigma$ sources expected to be due to random
fluctuations. Our sample contains 6 candidates at 5-$\sigma$ and 9 at
4-$\sigma$.

Our confidence about the robustness of the sample of bona fide
Lyman-$\alpha$ emitters is reinforced by a comparison with other
authors.  Steidel et al. (2000) focus on a known overdensity at
z$\simeq$3.09, with a fainter flux limit and a narrower filter. They
find a density of 0.9 sources \sq '$^{-1}$. Once this density is
reduced by a factor 6 to allow for their estimate of the field
overdensity, we derive a density of 0.16 sources \sq '$^{-1}$ which
compares favourably with our value of 0.07 sources \sq '$^{-1}$.  Out
of 77 candidates selected from the excess emission in the narrow band
filter, \cite{S2000} identify 5 as contaminants, i.e., a fraction of
6.4 \%. For our sample we have a more conservative fraction of
interlopers of 35 \%. This level of agreement is conforting given the
differences between the two samples.  Searching on a known
overdensity, Campos et al. (1999) had a flux limit of 6 $\times$
10$^{-17}$ erg cm$^{-2}$ s$^{-1}$ and $\Delta z \simeq 0.1$. They
found a surface density of objects of about 0.33 $\sq '$ $^{-1}$ at
z$\simeq$2.5 which is not incompatible with our result.  Other
searches (De Propris et al. 1993; Thompson et al. 1995) sampled
smaller areas at shallower limits so that our results remain
compatible with their upper limits.

\section{The sample}

The total Lyman-$\alpha$ emission from our sample is $2.1\times
10^{-14}$ erg cm$^{-2}$ s$^{-1}$ which increases to $(3.1 \pm
0.4)\times 10^{-14}$ erg cm$^{-2}$ s$^{-1}$ when corrected for
completeness. The star formation rate can be estimated directly from
the Lyman-$\alpha$ flux following \cite{charlotfall}. In the absence
of attenuation and adopting H$_0$=65 km s$^{-1}$ Mpc$^{-1}$, we find a
star formation rate per unit volume of 0.008 M$\odot$ yr$^{-1}$
Mpc$^{-3}$ for $\Omega =1$ or 0.005 M$\odot$ yr$^{-1}$ Mpc$^{-3}$ for
$\Omega_M=0.3$ $\Omega_\Lambda=0.7$. These values are accurate to
within a factor of 2. By comparing the measured value of the EW to
the theoretical prediction one obtains a direct measure of the
resonant attenuation of Lyman-$\alpha$.  Our sources have
$\overline{\rm EW}\simeq 39$, implying an attenuation correction of
$\sim3.8$. The corrected star formation rates per unit volume is 0.03
M$\odot$ yr$^{-1}$ Mpc$^{-3}$ for $\Omega =1$ or 0.02 M$\odot$
yr$^{-1}$ Mpc$^{-3}$ for $\Omega_M=0.3$ $\Omega_\Lambda=0.7$.  These
values are $\sim$17\% of the value determined by Madau et al. (1996)
for z$\simeq$ 2.75 and 5\% of the Lilly et al. (1996) peak at z=1.

A more empirical estimate can be obtained by adopting case B
recombination theory and the calibration by \cite{K94}. This would
yield about 15 \% of the value determined by Madau et al. (1996) for
z$\simeq$ 2.75 and 4\% of the Lilly et al. (1996) peak at z=1.  Higher
values would be estimated by following \cite{Giava} and adopting an
average Lyman-$\alpha$ attenuation by a factor $\sim$10.

An independent estimate of the SFR for our sample can be made on the
basis of their rest frame UV flux as derived from the B band
magnitudes. This will be uncertain to about a factor two because our B
filter includes also wavelengths beyond Lyman-$\alpha$ where
absorption by the Lyman-$\alpha$ forest can play a role. The same
calibration as in Madau et al. (1996) gives a star formation rate
$\sim$18 \% of the $z=2.75$ value, in good agreement with the
estimate based on the Lyman-$\alpha$ emission.

In Figure 2 we plot the EW vs the measured B-I
color for our sample (filled circles). The symbol size identifies
objects detected at 5 (largest), 4 (intermediate), and 3-$\sigma$
(smallest).  For reference we plot as open circles the points that are
considered as possible low redshift contaminants ([OII] emitters). The
dotted and dashed lines give the relation between EW and B-I color
derived on the basis of Bruzual \& Charlot (1993, BC96) models for
constant star formation and single bursts, respectively.  The model
spectra were corrected for the intergalactic attenuation due to
neutral hydrogen along the line of sight following the prescriptions
by Madau (1995). Clearly the red colors of our sources are compatible
with those observed by Pascarelle et al. (1998) in a field centered on
a known overdensity at z$\simeq$ 2.4 (open triangles) and in three
random WFPC2 fields (open stars). The observed colors and
Lyman-$\alpha$ EWs are incompatible with those of simple stellar
population models.  This was noted also by Campos et al. (1999) for
their sample.  A couple of Lyman-$\alpha$ emitting objects with very
red colors have also been studied by Francis et al. (2001).  A simple
explanation for the red colors is that these objects are heavily
reddened. However, this model requires some extreme fine tuning so
that Lyman-$\alpha$ can escape with relatively large EW while the I band
flux is dominanted by the reddened component.

An alternative explanation is that an older stellar population is
present in the sample objects. This is illustrated by the solid lines
in Figure 2 which show the resulting colors and equivalent widths
obtained by combining young populations with age between 1 and 10 Myrs
(top to bottom on each line) with an older stellar population 1 Gyr in
age. The mass ratio of the ``old'' to the young population is used to
label the curves.

Both explanations require the underlying ``hidden'' population to
greatly exceed the mass of the young population directly responsible
for the Lyman-$\alpha$ emission. For a two population mix with an age
of 1 Gyr for the oldest population, one needs ``hidden'' masses that
are 4000 and 8000 times the young population mass in order to obtain
B-I colors of 1.5 or 2, respectively. If the observed colors are due
to dust, one needs a reddened component 10000 times more massive than
the unreddened population. This factor has been calculated for the
case of a screen of dust. Other dust geometries would increase
extinction without increasing reddening and would lead to even larger
factors. The minimum masses quoted above have been determined by
considering a range from 100 Myr to 2 Gyr in age for the older
population in the population mix model, and dust reddening values
ranging from $A_V$ of 1 through 10 for the dust reddening model. We
have also found that our results do not depend critically on
metallicity.

\section{Discussion and Conclusions}

\label{sec:conclu}

By using a custom medium band filter, we have detected a large number
of bona fide Lyman-$\alpha$ emitters. A total of 31 blue candidates
have been excluded by our [OII] rejection criteria. However, their
inclusion would not have changed our conclusion that the majority of
our candidates are red. On average the emitters have colors much
redder than those expected for young, non-reddened,
starbursts. Objects with similarly red colors would be hard to find in
typical Lyman-break searches because such searches typically require a
blue color longwards of the break. The red color of Lyman-$\alpha$
emitters at z$\sim$ 2.4 was noted also by Campos et al. (1999) who,
however, did not attempt to model the effect. In order to account for
the red colors one needs a large fraction of the stellar mass to be
either old or highly reddened. Francis et al. (2001) reached a similar
conclusion for the two red compact objects they found within a more
extended Lyman-$\alpha$ nebula at z$\simeq$2.38. Note that this
conclusion (and the stellar masses derived below) would remain valid
even if we were simply detecting Seyfert nuclei rather than star
formation.  We do not think that this is the case since, by assuming
the Seyfert nuclei luminosity function at z=2.4 to be the local one
(e.g. \cite{seyfert}) scaled up to match that of QSOs at z$\simeq$2.3
(\cite{boyle}), we still find that less than a dozen of our final
sources would be expected to be an AGN.

If the observed colors are interpreted as due to dust reddening, the
global star formation rate from this population would have to be
enhanced by a factor 10000 or more, thus exceeding by two order of
magnitudes the global peak star formation rate estimated by
Madau et al. (1996). This makes it unlikely that reddening is the
general explanation of the red colors in our sample objects.

If instead an old population is present, one can estimate its relative
importance as follows. The minimum excess of old population is
obtained for a 1 Gyr old population. For the young population an age
of 5 Myrs is representative of the result (ages below about 10 Myrs
are needed to produce the observed EW). Assuming that the bursts are
separated by no less than 10 Myrs, we obtain a maximum of 100 for the
number of bursts that can have taken place. Since the mass of the old
population exceeds that of the young population by much more than a
factor 100, the old population must dominate the mass by at least a
factor 10, implying that the star formation episodes that we are
witnessing are neither the first nor the most important ones in the
life of these galaxies.  Indeed for a typical object in our sample, by
using Bruzual \& Charlot models and a Salpeter mass function between
0.1 and 125 M$_\odot$, we derive from the Lyman-$\alpha$ flux that the
mass of the unreddened young population is $\sim10^7$M$_\odot$,
implying total stellar masses of $\sim10^{11}$M$_\odot$. The number
density of our sources is about 5 per cent that of L$_\star$ galaxies
or brighter and is compatible with that of EROs or elliptical galaxies
as long as the duty cycle of accretion/star formation is high (50-100
per cent). These arguments suggest that these objects might have
formed already most of their stars by $z\simeq 2.4$ and therefore that
they might be proto-elliptical galaxies (\cite{spinrad},
\cite{stiavelli}).

\acknowledgments 

We thank S. Lilly and M. Fall for extensive discussions and
constructive criticism, the anonymous referee for suggestions that
helped improving the paper, M. Dickinson, R. Ellis and S. Djorgovski
for useful comments and P. Madau for making available his
intergalactic extinction curves. This project has been partially
supported by the SNS in Pisa, the MURST of Italy and by the STScI DDRF
grant 82241.

\clearpage

\clearpage
\begin{figure}
\plotone{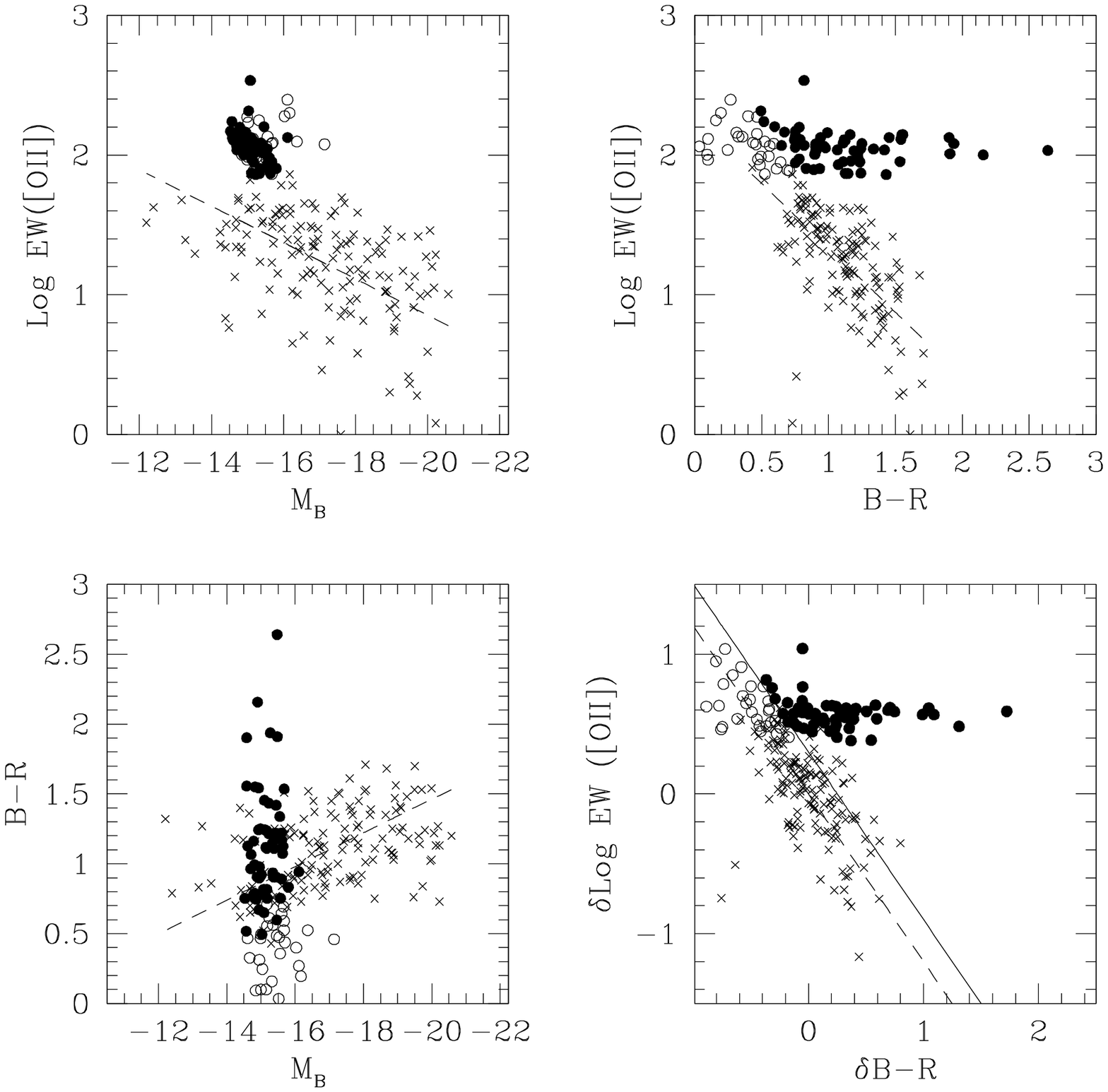}
\caption{We plot various correlations for local
[OII] emitters  (Jansen et al.\ 2000, crosses) and for our candidates
(open and filled circles) assuming all of our candidates are
actually interlopers located at z$\simeq0.11$.  On the Log EW
(observed) vs absolute magnitude plot (top left panel) the local [OII]
emitters are on average brighter and with lower EW than our
candidates. Similarly in the Log EW (observed) vs B-R color (top right
panel) the local [OII] emitters are bluer then our candidates. The
dashed lines are fits to the local [OII] emitters. The bottom left
panel shows the color-magnitude relation for the local [OII] emitters
and the lack of a correlation for our candidates. The bottom right
panel shows a delta-delta plot where for each object we have used the
EW vs M$_B$ and B-R vs M$_B$ fits to show the departures of EW and B-R
color from the expected values on the basis of the observed absolute
magnitude. This diagram is used by us to separate our bona fide
Lyman-$\alpha$ emitters (filled circles, to the right of the solid
line) from the possible [OII] contaminants (open circles, to the left
of the solid line).  }
\end{figure}

\begin{figure}
\plotone{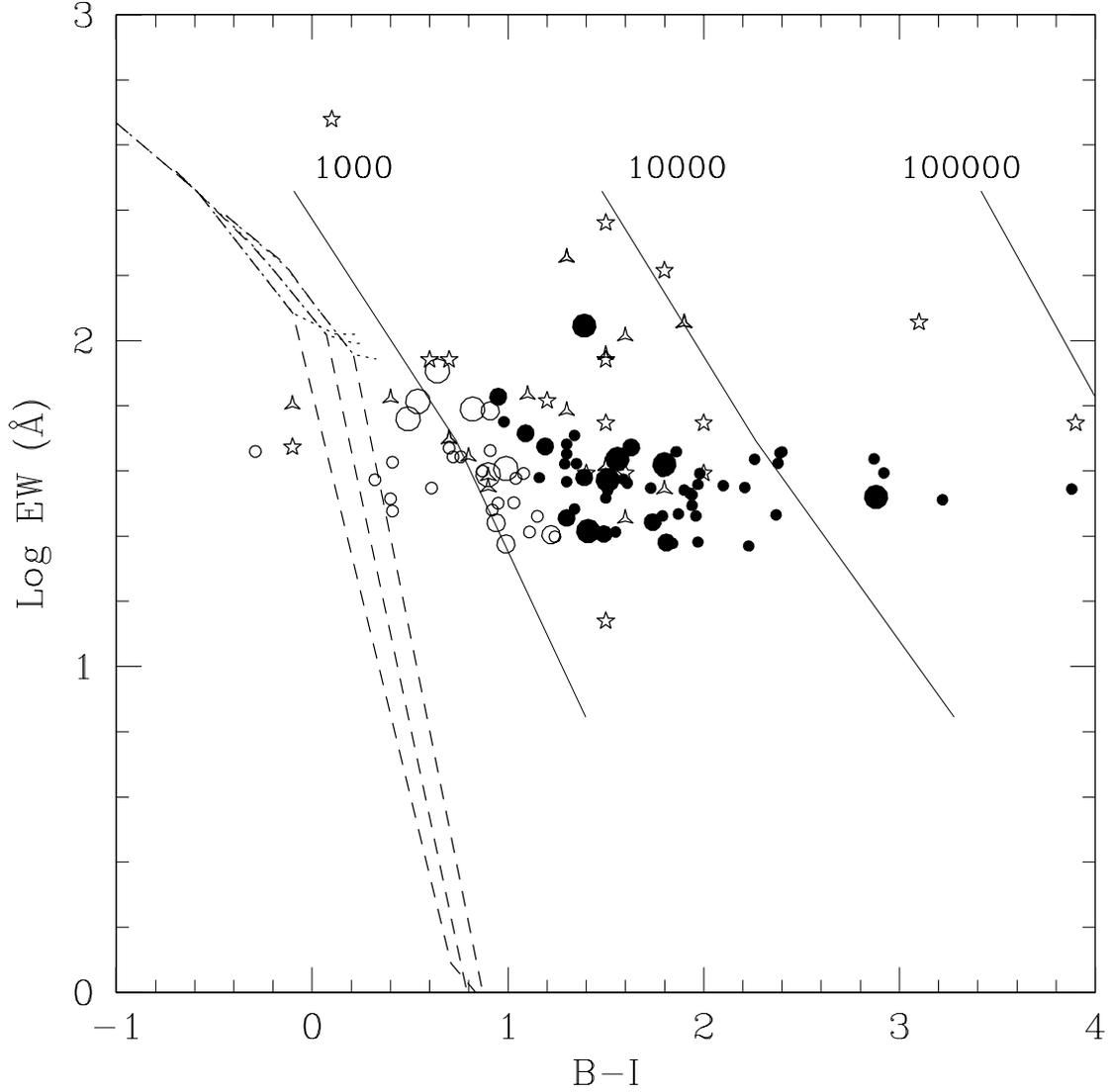}
\caption{The Log EW (rest frame) vs the observed B-I color is plotted
for our sample (filled circles).  The symbol size identifies objects
detected at a 5 (largest), 4 (intermediate), and 3-$\sigma$ (smallest)
level.  Possible low redshift contaminants ([OII] emitters) are
plotted as open circles. Also shown are the objects from Pascarelle et
al. 1998 for the cluster (open triangles) and the field (open stars).
The dotted and dashed lines give the relation between EW and B-I color
for continuous and single burst models, respectively, calculated using
Bruzual \& Charlot (BC96, with metallicity of 0.2 Z$_\odot$, 0.4
Z$_\odot$ and Z$_\odot$). The solid lines show the colors and
equivalent widths obtained by combining young populations with age
between 1 and 10 Myrs (top to bottom on each line) with an older
stellar population 1 Gyr in age. The mass ratio of the old to the
young population is used to label the curves. All the two component
models have metallicity 0.4 Z$_\odot$.}
\end{figure}


\begin{thebibliography}{}
%
\bibitem[Bertin \& Arnouts 1996]{SEx} Bertin, E., \& Arnouts, S. 1996,
\aaps, 117, 393
%
\bibitem[Boyle et al. 2000]{boyle}
Boyle, B.\ J., Shanks, T., Croom, S.\ M., Smith, R.\ J., Miller, L., Loaring, N., \& Heymans, C.\ 2000, \mnras, 317, 1014 
%
\bibitem[Bruzual \& Charlot 1993]{BC93} Bruzual, A.~G., \& Charlot, S.
1993, \apj, 405, 538
%
\bibitem[Burstein \& Heiles 1982]{maps} Burstein, D., \&  Heiles, C.,
1982, AJ, 87, 1165
%
\bibitem[Calzetti \& Heckman\ 1999]{Calzetti} Calzetti, D., Heckman, T.~M. 1999 \apj, 519, 27
%
\bibitem[Campos et al.\ 1999]{Campos} Campos, A., Yahil, A., 
Windhorst, R.\ A., Richards, E.\ A., Pascarelle, S., Impey, C., \& Petry, 
C.\ 1999, \apjl, 511, L1 
%
\bibitem[Carollo \& Lilly\ 2001]{LillyMarci} Carollo, C.~M., \& 
Lilly, S.~J., 2001, \apjl, 548, L153
%
\bibitem[Charlot \& Fall\ (1993)]{charlotfall} Charlot, S., \& Fall,
S.~M. 1993, \apj, 415, 580
%
\bibitem[De Propris et al.\ 1993]{DePropris} de Propris, R., Pritchet,
C.J., Hartwick, F.D.A., Hickson, P., 1993, \aj, 105, 1243
%
\bibitem[Francis et al. 2001]{francis} Francis,P.\ J., Williger,G.\ M., Collins, N.\ R., Palunas, P., Malumuth, E.\ M., Woodgate, B.\ E., Teplitz,H.\ I., 
Smette, A., Sutherland, R.\ S., Danks, A.\ C., Hill, R.\ S., Lindler, D.,
 Kimble, R.\ A., Heap, S.\ A., Hutchings, J.\ B. 2001, \apj, in press
%
%
\bibitem[Giavalisco, Koratkar, \& Calzetti\ (1996)]{Giava} 
Giavalisco, M., Koratkar, A., \& Calzetti, D.\ 1996, \apj, 466, 831 
%
\bibitem[Jansen et al.\ 2000]{Jansen} 
Jansen, R.\ A., Fabricant, D., Franx, M., \& Caldwell, N.\ 2000, \apjs, 
126, 331 
%
\bibitem[Kennicutt, Tamblyn, \& Congdon (1994)]{K94} Kennicutt, R.\ C., Tamblyn, P., \& Congdon, C.\ E.\ 1994, \apj, 435, 22 
%
\bibitem[Kennicutt 1998]{KenniLiege} Kennicutt, R. 1998, in The 
Next-Generation Space Telescope Science Drivers and Technological 
Challenges, ed. J. P. Swings \& B. Kaldeich-Schürmann (ESA SP-429; 
Noordwijk: ESA), 81
%
%
\bibitem[K{\"o}lher et al.\ 1997]{seyfert} Koehler, T., Groote, D., Reimers, D., \& Wisotzki, L.\ 1997, \aap, 325, 502 
%
\bibitem[Kobulnicky \& Koo\ 2000]{Kobulni} Kobulnicky, H.\ A.\ \& Koo, D.\ C.\ 2000, \apj, 545, 712
%
\bibitem[Landolt\ 1992]{Landolt} Landolt, A.~U. 1992, \aj, 104, 340
%
\bibitem[Lilly et al.\ 1996]{Lilly} Lilly, S.~J.,Le Fevre, O., Hammer, F., Crampton, D.\ 1996, \apj, 460, 1
%
\bibitem[McQuade, Calzetti \& Kinney 1995]{mcquade} McQuade, K., Calzetti, D.,
Kinney, A.L., 1995, \apjs, 97, 331
%
%
\bibitem[Madau 1995]{MP1} Madau, P. 1995, \apj, 441, 18
%
\bibitem[Madau et al.\ 1996]{MP} Madau, P., Ferguson, H.C., Dickinson, M.E.,
 Giavalisco, M., Steidel, C.C., Fruchter, A. 1996, \mnras, 283, 1388
%
\bibitem[Metcalfe et al.\ 2001]{Metcalfe} Metcalfe, N., Shanks, T.,
Campos, A., McCracken, H.J., Fong, R., \mnras, 323, 795
%
\bibitem[Meurer, Heckman, \& Calzetti 1999]{MHC} Meurer, G.R.,
Heckman, T.M., Calzetti, D. 1999, \apj, 521, 64
%
\bibitem[Moorwood et al.\ 2000]{Moorwood} Moorwood, A.F.M., van der Werf, P.P.,
Cuby, J. G., Oliva, E. 2000, \aap, 362, 9
%
\bibitem[Oke, 1974]{Oke} Oke, J.~B., 1974, \apjs, 27, 21
%
\bibitem[Pei \& Fall 1995]{PF95} Pei, Y.C. \& Fall, S.M. 1995, \apj, 454, 69
%
\bibitem[Pei et al.\ 1999]{Peietal} Pei, Y.C., Fall, S.M., Hauser, M.G. 1999,
\apj, 522, 604
%
\bibitem[Pettini et al.\ 2001]{Pettini2001} Pettini, M., Shapley,
A.E., Steidel, C.C., Cuby, J.-G., Dickinson, M., Moorwood, A.F.M.,
Adelberger, K.L., Giavalisco, M., 2001, ApJ, 554, in press
%
\bibitem[Schlegel et al. 1998]{EXMAPS} Schlegel, D.~J. et al.\ 1998, \apj, 500, 525
%
\bibitem[Scarlata et al.\ 2001]{Scarlata} Scarlata, C., Stiavelli, M.,
Lilly, S., Treu, T., Bertin, G., 2001, in preparation
%
\bibitem[Spinrad et al.\ 1997]{spinrad} Spinrad, H., Dey, A., Stern, D., 
Dunlop, J., Peacock, J., Jimenez, R., Windhorst, R. 1997, ApJ, 484, 581
%
\bibitem[Steidel et al.\ 1996]{Steidel} Steidel, C.\ C., 
Giavalisco, M., Pettini, M., Dickinson, M., \& Adelberger, K.\ L.\ 1996, 
\apjl, 462, L17 
%
\bibitem[Steidel et al.\ 1998]{Steidelb} Steidel, C.\ C., Adelberger, K.\ L.,
Dickinson, M., Giavalisco, M., Pettini, M., Kellogg, M. 1998, \apj, 492, 428
%
\bibitem[Steidel et al. (2000)]{S2000} Steidel, C.\ C., Adelberger, K.\ L., 
Shapley, A.\ E., Pettini, M., Dickinson, M., \& Giavalisco, M.\ 2000, 
\apj, 532, 170 
%
\bibitem[Stiavelli 1998]{StiavLiege} Stiavelli, M.  1998, in The 
Next-Generation Space Telescope Science Drivers and Technological 
Challenges, ed. J. P. Swings \& B. Kaldeich-Schürmann (ESA SP-429; 
Noordwijk: ESA), 71
%
\bibitem[Stiavelli et al. 1999]{stiavelli} Stiavelli, M., Treu, T.,
Carollo, C. M., Rosati, P., Viezzer, R., Casertano, S., Dickinson, M.,
Ferguson, H., Fruchter, A., Madau, P., Martin, C., Teplitz, H. 1999,
\aap, 343, L25
%
\bibitem[Thompson, Djorgovski \& Trauger 1995]{Thompson} Thompson, D.,
Djorgovski, S., Trauger, J., 1995, \aj, 110, 963
%
\bibitem[Treu \& Stiavelli 1999]{TT99} Treu, T., Stiavelli, M. 1999,
\apjl, 524, L27
\end{thebibliography}
\end{document}